# FROM DEEPFAKE TO DEEP-USEFUL: RISKS AND OPPORTUNITIES THROUGH A SYSTEMATIC LITERATURE REVIEW

Nikolaos Misirlis and Harris Bin Munawar
*HAN University of Applied Sciences, 6826CC, R31, Arhnem, The Netherlands*

**ABSTRACT**

Deepfake videos are defined as a resulting media from the synthesis of different persons' images and videos – mostly faces-, replacing a real one. The easy spread of such videos leads to elevate misinformation and represents a threat to society and democracy today. The present study aims to collect and analyze the relevant literature through a systematic procedure. We present 27 articles from scientific databases revealing threats for the society, democracies, the political life but presents as well advantages of this technology in entertainment, gaming, education, and public life. The research indicates high scientific interest in deepfake detection algorithms as well as the ethical aspect of such technology. This article covers the scientific gap since, to the best of our knowledge, this is the first systematic literature review in the field.

A discussion has already started among academics and practitioners concerning the spread of fake news. The next step of fake news considers the use of artificial intelligence and machine learning algorithms that create hyper-realistic videos, called deepfake. Deepfake technology is continuously attracting the attention of scholars the last 3 years more and more.

The importance on conducting research to this field derives from the necessity to understand the theory. The first contextual approach is related to the epistemological points-of-view of the concept. The second one is related to the phenomenological disadvantages of the field. Despite that, the authors will try to focus not only on the disadvantages of the field but also on the positive aspects of the technology.

**KEYWORDS**

Deepfake, Literature Review, Videos

## 1. INTRODUCTION

A discussion has already started among academics and practitioners concerning the spread of fake news (Durall et al., 2019, Korshunov and Marcel, 2018). The next step of fake news considers the use of artificial intelligence and machine learning algorithms that create hyper-realistic videos, called deepfake. Deepfake technology is continuously attracting the attention of scholars the last 3 years more and more.

Deepfake videos make their presence in almost every aspect of life. From politicians to celebrities, normal people, urban legend lovers and conspiracy theorists, all seem to be threatened or involved to the deepfake phenomenon. The vast majority of deepfake videos are related to pornography. A recent research from Deeptracelabs (deeptracelabs.com) indicated that 96% of deepfake videos are found to contain fake pornographic content with celebrities, attracting more that 134 million views. Only a smaller number of deepfake videos is related to politicians. The aforementioned research of Deeptrace found that more than 14000 deepfake videos were circulated online last year, respect to almost 8000 in 2018. The same research indicates that deepfake videos are not restricted to porn industry and politics, even if these two categories dominate the field. Together with the porn industry and the politics, criminals use artificial intelligence in order to impersonate CEOs voice and scam people (Stehouwer et al., 2019).

User-friendly apps render very easy to create and share deepfakes videos. Social media, especially content-sharing platforms, contribute to the ease in spreading these videos. Due to this fact several academics define the era we live a "post-truth" era, where the line between real and fake is extremely thin (Yatid, 2019, Neves et al., 2019, Sabir et al., 2019).

The study of deepfakes is important for both scholars and practitioners. Even though, scholarly research has only recently started to study the field with the literature to be still sparse (Öhman, 2019). The present research aims to cover this gap by reviewing the articles related only to deepfake from scholar databases. Previous research (Stehouwer et al., 2019)is conducted in the field, only by reviewing magazines' (non-academic) articles. The originality of this paper consist to the fact that this is the first systematic literature review for deepfake-only related articles from scientific-only databases.

The article is structured as follows. After the introduction, the methodology followed for the review is analyzed. Consequently, the major findings of the research are presented as well as the upcoming threats of the deepfake technology. Furthermore, the opportunities and the positive aspects from the use of deepfakes are discussed. The study concludes with future implications and limitations.

The importance on conducting research to this field derives from the necessity to understand the theory. The first contextual approach is related to the epistemological points-of-view of the concept. The second one is related to the phenomenological disadvantages of the field. Despite that, the authors will try to focus not only on the disadvantages of the field but also on the positive aspects of the technology.

## 2. MATERIAL AND METHODS

The Cochrane method is used for the systematic literature review and the findings are presented in accord with the PRISMA guidelines (Preferred Reporting Items for Systematic Reviews and Meta-Analyses). Google Scholar, EBSCO, Scopus and ScienceDirect were first searched, until January 2020, using a search strategy based on one single keyword and its possible combinations – Deepfake OR 'Deep fake'. The search strategy was then applied to each of the aforementioned databases, all articles were downloaded and merged into a single library in order to check for duplicate articles.

### 2.1 Results and Study Characteristics

The initial research returned 859 articles. After the duplicates removal, 825 articles were remained. From them, 784 articles were excluded due to several reasons, mostly irrelevance. The title of the articles were relevant to the topic discussed (deepfake), but after the first read, there was no relevance with our research.

Furthermore, articles from: Newspapers, citations, websites, blogs, vlogs, theses or dissertations, annual reports, white papers from companies, non-English journals, editorials, posters and presentations, eBooks, interviews and preprints were excluded from our research. The originality of our research is based on the fact that, to the best of our knowledge, this is the first literature review of scientific articles, therefore any other type of article should be excluded from the analysis. The topic is relatively new and academic search engines show results that should not be included to an academic literature review. As a consequence, from a vast amount of results (n=859), we finally are able to take into consideration only 3.1% of those. From the remaining 41 articles, 14 were excluded after the first screen due to irrelevance. Only 27 articles met the inclusion criteria (Figure 1).

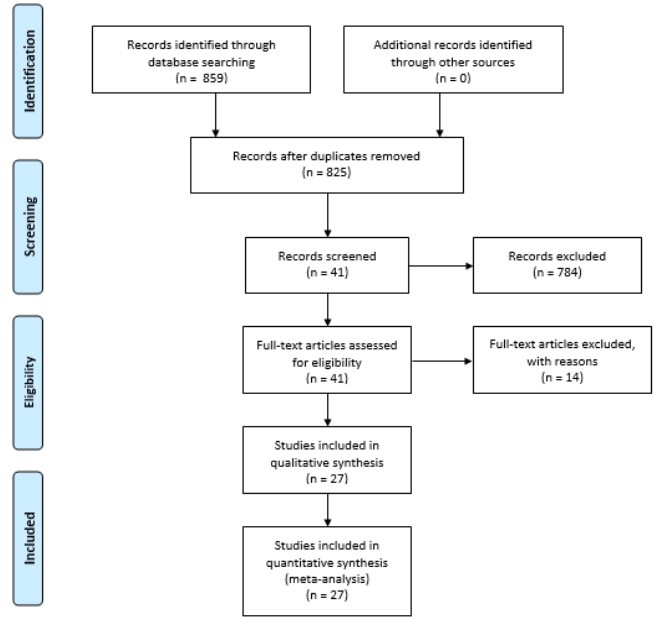

Figure 1. PRISMA flowchart depicting the study selection process

All 27 articles were published either in 2018 (n=5, 18.5%) or in 2019 (n=22, 81.5%). This metric by itself can be interpret as an index how much 'young' the topic is and explain the low number of publications. This field of science is in its infancy. Practitioners and R&Ds from companies are ahead, with Academia to follow in this case. This observation by itself enforces the fact that more research needs to be done, not only on the technological aspects of the field but also from a sociological – philosophical perspective. 59% of the articles come from American institutions (63% if we consider a UK/USA collaboration on an article). Other countries' institutions that contribute the research are Australia, Portugal together with Spain, Germany, United Arab Emirates and Taiwan with only one article for each country (3.7%) and Switzerland, The Netherlands, Italy with 2 articles (7.4%). Table 1 summarizes the 27 articles studied. For each article, the topics, the tools and the methods used are briefly explained.

Table 1. Topics covered by the articles of the review, including country of the authors' affiliation and year of publication

| Author(s) & Year | Country | Topics | Tools and/ or methodology* |
|---|---|---|---|
| Akhtar and Dasgupta (2019) | USA | Comparative experimental investigation & Face authenticity | Face manipulation detection |
| Diakopoulos and Johnson (2019) | USA | US 2020 electoral integrity and ethical issues | Scenarios development |
| Westling (2019) | USA | Emerging technologies and online platform regulations | - |
| Hasan and Salah (2019) | UAE | Deepfake detection. Use of Ethereum smart contracts to trace and track the provenance and history of digital content to its original source | Blockchain-based |
| Hsu et al. (2020) | Taiwan | Forgery detection and identification. | deep learning-based approach |
| Chesney and Citron (2018) | USA | Risks for democracy and national security | - |

| Author(s) & Year | Country | Topics | Tools and/ or methodology* |
|---|---|---|---|
| Kozemczak (2019) | USA | Risks for democracy | - |
| Nguyen et al. (2019) | Australia | Threats to privacy, democracy and national security | Literature review on deepfake detection |
| Amerini et al. (2019) | Italy | Adoption of optical flow fields to exploit possible inter-frame dissimilarities | Forensic technique able to discern between fake and original videos |
| Güera and Delp (2018) | USA | Use of convolutional neural network (CNN) to extract frame-level features | Train a recurrent neural network (RNN) |
| Korshunov and Marcel (2018) | Switzerland | Deepfake detection – tampering detection | Algorithms' evaluation |
| Dixon Jr (2019) | USA | Effects of deepfake in political campaigns, business interests and video evidence in court rooms | - |
| Koopman et al. (2018) | The Netherlands | Forensic challenges Detection of deepfake manipulation through | Photo Response non Uniformity (PRNU) model |
| Maras and Alexandrou (2019) | USA | Pornographic images analysis | - |
| Farish (2019) | UK, USA | UK's legislation analysis | - |
| Yang et al. (2019) | USA | Exposition of AI-generated fake face images or videos | Error revealing |
| Li et al. (2018) | USA | Eye-blinking detection for deepfake video detection | - |
| Öhman (2019) | UK | Ethical aspects of deepfake. Analysis of the pervert's dilemma (deepfake vs. private sexual fantasy) | - |
| Stehouwer et al. (2019) | USA | Detection of digital face manipulation | Use of databases with numerous types of facial forgeries |
| Agarwal et al. (2019) | USA | Threats to democracy & national security | Forensic technique for facial expressions and movements |
| Neves et al. (2019) | Portugal, Spain | Face-synthesis detection system. | Use of free datasets |
| Sabir et al. (2019) | USA | Deepfake detection | Faceforensics++ |
| Metaxas (2018) | USA | Ethical policies and epistemological education | Definition for deepfakes, classification of deepfake types and identification of risks and opportunities |
| Dolhansky et al. (2019) | USA | Deepfake video detection | DFDC datasets |
| Wagner and Blewer (2019) | USA | deepfake videos evaluation on enforcing gendered disparities within visual information | - |
| Durall et al. (2019) | Germany | Deepfake video detection | Classical frequency domain analysis. Image forensics and forgery detection |
| Korshunov and Marcel (2019) | Switzerland | Deepfake detection | VidTIMIT database |

* n/a tool or methodology when the topis is self-explanatory

## 3. RESULTS

### 3.1 Major Findings and Characteristics of the Articles

Regarding the topics in each article, deepfake detection articles dominate the literature with 15 papers to be focused completely on this topic. 6 articles are focusing on the risks and threats to society and democracy and 2 consider strictly the ethical aspects of the topic. Regarding the threats, the following 4 are the most discussed in research today, namely: politics, technological skepticism, outdated legal framework, use of celebrities, and not only, to pornographic-related videos and threats for the democracy and our society. The following paragraphs will sum up the findings of the research.

### 3.2 Threats and Risks

The speed of disinformation spread seems to be one of the worst threats of Democracy today (Agarwal et al., 2019, Chesney and Citron, 2018, Kozemczak, 2019, Nguyen et al., 2019). Social media facilitate this process by spreading information among users who prefer to share without first verifying the content (Dekker et al., 2020, Stover, 2018, Wirth, 2019). When deepfakes are used in order to criticize and satirize public figures or politicians, it's too obvious to understand the fake content. But the line between what is real and what can harm the society is still indistinct. It is under dispute whether deepfake technology will facilitate or not the overall structure of the society. An obvious satire on politicians and parodies could be also accepted, if unharmful, but what happens when deepfake videos affect the global stock market by creating rumors or even worst, when deepfakes involve terrorist actions?

Since deepfake videos represent a rather new technology, the legal legislation is still weak. To the best of our knowledge, California is the only state in US that already signed a legislation that makes distribution of digital material related to deepfake, illegal, 60 days before elections (TÉCNICO, 2014). Maybe, more states need to follow California's paradigm, especially now that the US elections approach. The latest research of Deeptracelabs indicates that 96% of deepfake videos concern pornography, but it is easy to misuse porn for political purposes. The case of Rana Ayyub is showing clearly that the distinction between politics and pornography is difficult to define.

Another threat of deepfake is related to citizens' trust towards information, technology, journalism and even democracy. Citizens are losing trust in institutions, becoming more and more tech-skeptic and apathetic - what's called *"information apocalypse" and "reality apathy"* (Stover, 2018, Wirth, 2019). As a result of this lack of confidence in the media, citizens will perceive as fake news even those cases which are true only because they are convinced that what did not fit their opinion must be fake.

Though expected, most of the articles focus on the 'dark' side of this technology. The present study, even though it recognizes how serious the misuse of technology can be, will analyze some positive aspects of deepfake and propose future implications of this technology. The reviewed literature focused almost exclusively on the 'dark' side of this technology and systematically ignored its positive, ethical or beneficial future applications and societal implications. This result was in line with the expectations of the authors of this study. The outcome is discussed in more detail in the next section.

## 4. DISCUSSION

The discourse in the reviewed literature almost entirely consists of the risks, threats and challenges of deepfake technology, and ignores some beneficial ways in which it can contribute to the fields of entertainment, education, healthcare and business. The authors argue that this indicates a bias. This is especially noticeable in our study because we excluded journalistic publications from the review. A comparable review by Westerlund (2019) has a section about the benefits of deepfake technology that exclusively sites non-academic publications.

Based on examples of specific applications of deepfake and adjacent technology in academic publications, the researchers will point out some of the potential beneficial uses of such technology ignored in present academic literature.

In the entertainment sector, applications of deepfake technology are not limited to pornography. Filmmakers can also benefit from this technology with realistic depictions of actors who are not available, or even dead, such as the brief appearance of Peter Cushing in the 2016 Star Wars film Rogue One, and James Dean being featured in the 2020 film Finding Jack (Monroe, 2020). The Dali Museum in Florida features a deepfake of the iconic surrealist painter to engage its visitors (Mihailova, 2021) and is a telling example of the application of deepfake technology for education, especially history education. These examples of virtual immortality also open doors for a critical look at identity and subjectivity.

In the health sector, research has shown that Augmented and Virtual Reality simulations have significant potential benefits in the areas of mental health and pain management (for example: Riva et al. (2019). A 2020 south Korean documentary showed the gratitude of a mother who was able to meet, in virtual reality, her daughter who had died four years prior (Stein, 2021). Deepfake technology can enhance virtual reality and thus improve such interventions (Bose and Aarabi, 2019).

In the business sector, as sales move online, returns have significant financial implications for businesses and have consequences relating to the environment and logistics (Cullinane et al., 2019). Virtual try-ons help reduce such returns (Hwangbo et al., 2020) but there are limitations in the use of virtual reality technology for this purpose (Boletsis and Karahasanovic, 2020). Advancements in deepfake technology can help improve the possibility of trying on clothing and cosmetics on your own virtual image before buying. This will eventually mean lesser need for human supermodels, who usually fall in a narrow range of body types and skin tones. The use of deepfake technology to make popular footballer David Beckham speak nine languages in the Malaria Must Die campaign indicates that its use in marketing is not necessarily deceptive (de Ruiter, 2021).

## 5. CONCLUSION

This is the first systematic review of deepfake-related articles publish only in scientific journals. The research revealed only 27 relevant articles that fit the field of study, due to the fact that the field itself is in its infancy. All articles are related to the negative aspect of the deepfake technology, though there is still ground to pave and look for opportunities. The present literature review present a solid base for future researchers to build more complete review in the years to come. Together with this, the authors of the present study want to focus on the positive implications of deepfake technology in society that will be formed from more responsible and active citizens, in order to enforce even more the democracy around the globe. Deepfakes are only a tool, like knifes are. It is up to citizens, scientists and practitioners to transform this knife from a deadly weapon to a scalpel that will benefit the society.